# Bots, #StrongerIn, and #Brexit: Computational Propaganda during the UK-EU Referendum

**Philip N. Howard**
Oxford University
philip.howard@oii.ox.ac.uk
@pnhoward

**Bence Kollanyi**
Corvinus University
kollanyi@gmail.com
@bencekollanyi

**ABSTRACT**
*Bots are social media accounts that automate interaction with other users, and they are active on the StrongerIn-Brexit conversation happening over Twitter. These automated scripts generate content through these platforms and then interact with people. Political bots are automated accounts that are particularly active on public policy issues, elections, and political crises. In this preliminary study on the use of political bots during the UK referendum on EU membership, we analyze the tweeting patterns for both human users and bots. We find that political bots have a small but strategic role in the referendum conversations: (1) the family of hashtags associated with the argument for leaving the EU dominates, (2) different perspectives on the issue utilize different levels of automation, and (3) less than 1 percent of sampled accounts generate almost a third of all the messages.*

**FROM SOCIAL BOTS TO POLITICAL BOTS**
A growing number of political actors and governments worldwide are employing both people and bots to engage in political conversations online.[1]–[3] Bots can perform tasks that range from legitimate, like generating a large amount of benign tweets that deliver news or update feeds, to more malicious, like spreading spam by delivering appealing text content with the link-directed malicious content. Whatever their uses, all bots share the property of being able to deploy messages and replicate themselves.[4]

Networks of such bots are called "botnets," a term combining "robot" with "networks" and describing a collection of connected computers with programs that communicate across multiple devices to perform some task. There are legitimate botnets, like the Carna botnet, which gave us our first real census of device networks,[5] and there are malicious botnets, like those that are created to launch spam and distributed denial-of-service (DDoS) attacks and to engineer theft of confidential information, click fraud, cyber-sabotage, and cyberwarfare.

Social bots are particularly prevalent on Twitter, but they are found on many different social media platforms that increasingly form part of the system of political communication in many countries.[6] They are computer-generated programs that post, tweet, or message of their own accord. Often bot profiles lack basic account information such as screen names or profile pictures. Such accounts have become known as "Twitter eggs" because the default profile picture on the social media site is of an egg. While social media users get access from front-end websites, bots get access to such websites directly through a code-to-code connection, mainly through the site's wide-open application programming interface (API) that enables real-time posting and parsing of information.

Bots are versatile, cheap to produce, and ever evolving. "These bots," argues Rob Dubbin, "whose DNA can be written in almost any modern programming language, live on cloud servers, which never go dark and grow cheaper by day."[7] Unscrupulous internet users now deploy bots beyond mundane commercial tasks like spamming or scraping sites like eBay for bargains. Bots are the primary applications used in carrying out distributed denial-of-service and virus attacks, email harvesting, and content theft. A subset of social bots are given overtly political tasks and the use of political bots varies across regime types. Political actors and governments worldwide have begun using bots to manipulate public opinion, choke off debate, and muddy political issues. Political bots tend to be developed and deployed in sensitive political moments when public opinion is polarized. How have bots been used in the political conversations about the UK's role in the EU?

**TWITTER BOTS AND POLITICAL COMMUNICATION**
Twitter has become a powerful communication tool during many kinds of crises, political or otherwise. When drug wars reemerged in Mexico recently, neither the drug lords nor the government expected a network of real-time war correspondents to spring up to report battles between police and gangs. Tweeting certainly didn't stop the drug war. But it helped people to cope. We can't measure how important the sense of online community provided by active social media use can be when modern voters discuss politics. And while the quality of deliberations over social media can appear base, social media is one mode of several modes of deliberation that contemporary voters have. Over social media like Twitter, a few citizens often rise to the occasion, curating content and helping to distinguish good information from bad.[4, p. 22], [8]

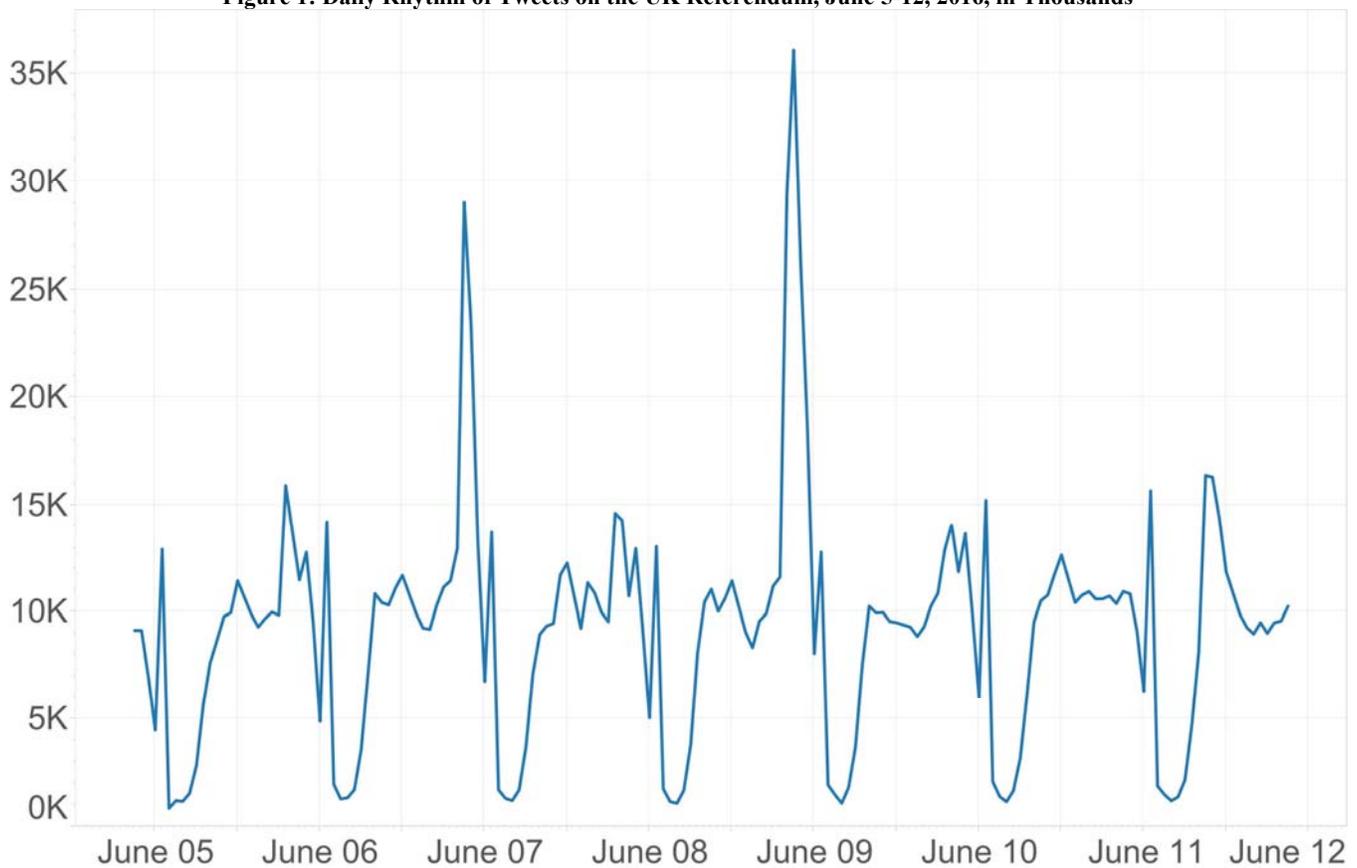

Figure 1: Daily Rhythm of Tweets on the UK Referendum, June 5-12, 2016, in Thousands

*Source: Author's calculations based on Twitter sample of these hashtags June 5-12, 2016.*
*Note: This table reports the number of Tweets, hour by hour, during the sample period.*

In the case of the StrongerIn-Brexit debate, the two single most active accounts from each side of the debate are bots. These accounts, @ivoteLeave and @ivotestay, follow a similar algorithm. Neither generates new content, they mechanically retweet messages from their side of the debate. Another surprise bot is @Col_Connaughton, a long-running pro-Palestine bot that seems to have been repurposed to support the Brexit side. It is not clear if there is any human curation involved, but it certainly uses automation. Another pro-Palestinain bot is @Rotenyahu, which has been tasked to retweet messages from Col_Connaughton account so it too is distributing #Brexit content. These are good examples of bots designed to amplify a source simply by aggregating and repeating content.

**SAMPLING AND METHOD**
This data set contains more than 1.5 million Tweets collected June 5-12, 2016, using a combination of pro-leave, pro-remain and neutral hashtags to collect the data. This sampling strategy yielded 313,832 distinct Twitter user accounts. Since our purpose is to discern how bots are being used to amplify political communication on this important policy question, we did some basic descriptive analysis to understand the rhythm of social media activity on this topic.

The flow of bot traffic varies during political crises. But for our sample period, Figure 1 presents the daily rhythm of Tweets on the UK referendum. Given the limits that Twitter places on researchers, it is impossible to report the total number of bots engaged in the StronterIn-Brexit debate.[10]

Twitter provides free access to a sample of the public Tweets posted on the platform. The exact sampling method is not known, but according to Twitter, the data available through the Streaming API is at most one percent of the overall global public communication on the platform at any given time.[9] In order to get the most complete and relevant data set, the tweets were collected by following a list of hashtags. The list of hashtags was generated by downloading two smaller data sets and extracting the hashtags that were most often used on both sides of the Brexit debate as well as some of the neutral hashtags used to refer to the EU referendum. The programming of the data collection and most of the analysis were done by using the statistics package R.

Selecting tweets on the basis of hashtags has the advantage of capturing the content most likely to be about this important political issue (and excluding, in our experience, Eurovision-related tweets). The streaming API yields (1) tweets which contain the keyword or the hashtag; (2) tweets

Table 1: Hashtag Use on Twitter, by Perspective on the UK Referendum

| Perspective | N | % |
|---|---|---|
| Remain (#strongerin, #remain, #voteremain, #votein, #bremain, #labourin, #votestay, #intogether, #labourinforbritain, #greenerin) | 363,217 | 20 |
| Leave (#brexit, #voteleave, #leaveeu, #takecontrol, #betteroffout, #voteout, #beleave, #brexitthemovie, #euistheproblem, #brexitbustour) | 993,176 | 54 |
| Neutral (#euref, #eureferendum, #inorout, #eudebate, #june23) | 475,233 | 26 |
| Occurrence of All Above Hashtags | 1,831,626 | 100 |

*Source: Author's calculations based on Twitter sample of these hashtags June 5-12, 2016.*
*Note: This table reports the number of times these hashtags were used, not the number of tweets.*

with a link to a web source, such as a news article, where the URL or the title of the web source includes the keyword or hashtag; (3) retweets where the text contains the original text, and the keyword or hashtag is used either in the retweet part or in the original tweet; and (4) retweets where the original text is not included but Twitter uses a URL to refer to the original tweet. From our data set, for example, appears this tweet:

*"Ask Sam to get your stuff packed then, @David_Cameron https://t.co/ysoqi9MoQ0"*

It uses no hashtag, but the URL leads to the original tweet that uses #voteleave and #takecontrol.

**FINDINGS AND ANALYSIS**

With this sample, we can draw some conclusions about the character and process of political conversation over Twitter on this topic. Specifically, we can parse out the amount of social media content related to the various positions in the debate, and we can investigate how much of this content is driven by bots. We can parse out the volume of tweets by perspective, assess the level of automation behind the different perspectives, and evaluate the particular contribution of bots to traffic on this issue.

First, to evaluate the social media content on Twitter about the UK referendum we can do some simple analysis of the frequency that particular hashtags are used by users—or bots—to signal their perspective and locate their post in larger conversations. Table 1 reveals the frequency of use for ten hashtags that are usually associated with #StrongerIn arguments, ten that are usually associated with #brexit arguments, and five that are relatively neutral.

Our method counted tweets with selected hashtags in a simple manner. Each tweet was coded and counted if it contained one of 25 specific hashtags that were being followed. If the same hashtag was used multiple times in a tweet, this method still counted that tweet only once. If a tweet contained more than one selected hashtag, it was credited to all the relevant hashtag categories.

Analyzing sentiment on social media such as Twitter is difficult.[11], [12] Obviously, contributions using none of these hashtags are not captured in this data set. Tweets that use multiple hashtags are counted multiple times, and it is not uncommon for users to mix the use of hashtags. Each occurrence of a hashtag is counted separately, so one tweet with multiple hashtags could be in more than one group. These counts only include tweets where the text of the tweet contains one of the hashtags. This is the majority of the tweets, but it excludes the newer type of retweets in which a user comments and the original tweet is retweeted in the form of a URL.

Understanding how #brexit is used illustrates how complex hashtag use is. It is a tricky hashtag because it has often been used to support both sides of the argument. For comparison, two recent tweets using the same hashtag encourage voters to do two different things:

*"If you're unsure how to vote in the UK Referendum just look at how Murdoch is voting & do the opposite. Vote to stay in the Union #brexit"*

*"What #Brexit Could Mean for Travelers #UK #Ireland #Europe #Tourism http://www.tourismmarketer.com/articles/europe/uk-and-ireland/what-brexit-could-mean-for-travelers-28352-thread.html..."*

These are exceptions—most instances of #brexit still accompany messages about leaving the EU. The second most popular #brexit hashtag (#voteleave) is still used three times more than the most frequent encouraging membership (#StrongerIn). The hashtag #voteleave appears 341,839 times in the data set, while #StrongerIn appears 110,653 times. But it is still worth noting the contrast, and computational social scientists do not yet understand the sampling parameters sufficiently to make many inferences about how opinion on social media translates to voter intent.[13]

**Table 2: Twitter Content, by Hashtag Use and Level of Bot Automation**

|  | All Tweets | | Generated With Heavy Automation | | Generated By Disclosed Bots | |
|---|---:|---:|---:|---:|---:|---:|
|  | N | % | N | % | N | % |
| Exclusively StrongerIn (number of tweets that used one or more of only #strongerin hashtags) | 186,279 | 14.6 | 28,075 | 15.1 | 196 | 0.1 |
| Exclusively Brexit (number of tweets that used one or more of only #brexit hashtags) | 662,745 | 51.8 | 97,431 | 14.7 | 842 | 0.1 |
| Exclusively Neutral (number of tweets that used one or more of only Neutral hashtags) | 234,170 | 18.3 | 13,436 | 5.7 | 253 | 0.1 |
| Mixed, Brexit-Neutral | 69,322 | 5.4 | 11,667 | 16.8 | 72 | 0.1 |
| Mixed, StrongerIn-Neutral | 35,412 | 2.8 | 5,099 | 14.4 | 44 | 0.1 |
| Mixed, Brexit-StrongerIn | 49,556 | 3.9 | 9,735 | 19.6 | 89 | 0.2 |
| Mixed, Brexit-StrongerIn-Neutral | 40,926 | 3.2 | 13,640 | 33.3 | 35 | 0.1 |
| Total | 1,278,410 | 100.0 | 179,083 | 14.0 | 1,531 | 0.8 |

*Source: Author's calculations based on Twitter sample of these hashtags June 5-12, 2016.*
*Note: Heavy automation refers to tweets generated by accounts that produce more than 50 tweets per day. Disclosed bots either use the term "bot" in the Twitter handle or make use of a known bot launching platform.*

Unfortunately, not enough users geotag their profiles to allow analysis of the distribution of this support around the world or within Europe. Nonetheless, this data does reveal that users tweeting from the Brexit perspective (a) have generated a larger block of content and (b) are better at tagging their contributions so as to link messages to a larger argument and wider community of support.

Second, to evaluate the role of bots in this debate, we organize clusters of opinion based on hashtag use. Table 2 distinguishes between the messages that exclusively used a hashtag known to be associated with a perspective and then the combinations of mixed tagging that are possible. Then we create a subcategory of accounts that use heavy levels of automation.

These accounts are often bots that see occasional human curation, or they are actively maintained by people who employ scheduling algorithms and other applications for automating social media communication. We define "heavy automation" as accounts that post at least 50 times a day, meaning 350 or more tweets during the data collection period. Extremely heavy human users might achieve this pace of social activity, especially if they are simply retweeting the content they find in their social media feed. And some bots may be relatively dormant, waiting to be activated and tweeting only occasionally. But this threshold generally captures accounts generating large traffic with some level of automation. Finally, self-disclosed bots were identified by searching for the term "bot" in either the tag or account description. While this is a small proportion of the overall accounts, we expect the actual number of bots to be higher—many bots, after all, would not disclose their activities. Future research will involve a more detailed analysis of the disclosed and hidden bots and searching for a wider range of terms referring to bots in the account name and description data.

Third, to understand the distribution of content production across these users, we then look at segments of the total population of contributors to these hashtags. We find that the top 100 users generated more than 121,000 tweets during the week, which is about 8 percent of all Twitter traffic related to StrongerIn-Brexit.

The most active users—the accounts that tweeted 100 or more times with a related hashtag during the week—generated 32 percent of all Twitter traffic about Brexit. That volume is significant, considering that this number of posts was generated by fewer than 2,000 users in a collection of more than 300,000 users. In other words, less than 1 percent of the accounts generate almost a third of all the content. However, not all of these users or even the majority of them are bots. Anecdotally, it is difficult for human users to maintain this rapid pace of Twitter activity without some level of account automation.

This table reveals that automation is used at several different levels by people taking different perspectives in the debate. The accounts using exclusively neutral hashtags are rarely automated (only 5.7 percent use heavy automation) while one-third of all the tweets using a mixture of all hashtags are generated by accounts that use heavy automation. Yet, surprisingly, only a fractional number of accounts disclose that they are in fact bots.

Third, to evaluate the role of bots in generating traffic on StrongerIn-Brexit topics, we took a close look at the top 10 accounts by volume, and all of them seem to use some level of automation. It is almost certain that 7 of the 10 accounts are bots. One of them is a UKIP-curated account most probably with some level of automation. Two of them seem to be bots that get some small amount of human curation. On the whole, these top users do not generate new content but simply retweet content from other users.

Keeping track of bots—especially political bots—requires careful understanding of how the design features of platforms may constrain the sampling strategy. Rather than trying to determine which specific tweet was generated by a bot, we look at the type of platform used to retweet or create a tweet to determine the probability that it was bot-generated. Many of the accounts that are driven by a bot use bot-dominated platforms and normal platforms for tweeting activity. Indeed, some of the accounts that we identified as likely being bots have since been suspended by Twitter—the company also considered them to be bots.

**CONCLUSIONS**

We tend to think of the Internet in general, and social networks in particular, as connecting human beings. And it's true that the Internet permits us to connect and convene at an unprecedented scale. But the Internet is also famously mediated. We do not reach one another directly so much as through a layer of technology—an interface, a platform, a network—that someone else has designed. What this means in part is that some of the personalities we encounter in cyberspace are not who or what they purport to be. In fact, people are increasingly agreeing, arguing, and even flirting with fleeting bits of code known as bots.

It is no secret that citizens, journalists, and political leaders now make use of political bots—automated scripts that produce content and mimic real users. But it is not clear that average users can distinguish bot from human activity. Political campaigns are complex exercises in the creation, transmission, and mutation of significant political symbols.[14] Increasingly, political campaigns automate their messaging and many citizens who use social media are not always able to evaluate the sources of a message or critically assess the forcefulness of an argument. Fake social media accounts now spread pro-governmental messages, beef up website follower numbers, and cause artificial trends. Political strategists worldwide are using bot-generated propaganda and misdirection. Research suggests that when digital media become an important part of civic engagement, social movements can generate immense amounts of content that cascade across public conversations—both across social media platforms and over international borders.[15] And increasingly, political elites have been learning and applying communication innovations by activists as tools for social control.

The measures of undecided voters suggest that 30 percent of UK voters will decide how to vote in the week before the election, and half of these will decide on polling day.[16] The pervasive use of bots over social media heightens the risk of massive cascades of misinformation at a time when voters will be thinking about their options and canvasing their social networks for the sentiments of friends and family.

Bots have been used by political actors around the world to attack opponents, choke off hashtags, and promote political platforms. During this sample period, however, we found that social media bots were used mostly for amplifying messages rather than argumentative engagement or even impression management.

Robotic lobbying tactics have been deployed in many countries, including Russia, Mexico, China, Australia, the United Kingdom, the United States, Azerbaijan, Iran, Bahrain, South Korea, Turkey, Saudi Arabia, and Morocco. Indeed, experts estimate that bot traffic now makes up over 60 percent of all traffic online—up nearly 20 percent from two years prior.[17]

Bots operate on many sensitive political topics during close electoral contests in many advanced democracies.[3], [18], [19] Political algorithms have become a powerful means of political communication for "astroturfing" movements—defined as the managed perception of grassroots support.[20] In this way bots have become a means of managing citizens, They have gone from simply padding follower lists to retweeting volumes of their own commentary and announcements. In this analysis, we find that bots generate a noticeable portion of all the traffic about the UK referendum, very little of it original. Repeating this sample collection and study method over a longer time period and right around voting day would likely reveal additional features of bot activity on this topic.

Bots are mostly used to retweet content about StrongerIn-Brexit issues. It is difficult to say how much public opinion is shaped by political discourse on this topic over social media or what the influence of bots is on public sentiment. Nevertheless, we can identify the role that bot algorithms play in political communication around StrongerIn-Brexit issues. We find that political bots have a small but strategic role in the referendum conversations: (1) the family of hashtags associated with the argument for leaving the EU dominates, (2) different perspectives on the issue utilize different levels of automation, and (3) less than 1 percent of sampled accounts generate almost a third of all the messages.

**ABOUT THE TEAM**

**Philip N. Howard** is a Statutory Professor in the Oxford Internet Institute at the University of Oxford. He has published eight books and over 100 academic articles, book chapters, conference papers, and commentary essays on information technology, international affairs and public life. Howard is the author of *The Managed Citizen* (Cambridge, 2006), the *Digital Origins of Dictatorship and Democracy* (Oxford, 2010), and most recently of *Pax Technica: How the Internet of Things May Set Us Free or Lock Us Up* (Yale, 2015). He blogs at www.philhoward.org and tweets from @pnhoward.

**Bence Kollanyi** is a PhD student in Sociology at the Corvinus University of Budapest currently living in Copenhagen, Denmark. He is also studying political bots in the Computational Propaganda project at the Oxford Internet Institute. His research focuses on automation of social media accounts, including the development of open-source bots, and their deployment on Twitter and other social media platforms. He tweets from @bencekollanyi.

**ABOUT THE PROJECT**

The Project on Computational Propaganda (www.politicalbots.org) involves an international, interdisciplinary team of researchers investigating the impact of automated scripts—computational propaganda—on public life. We track social bots and use perspectives from organizational sociology, human computer interaction, communication, and political science to interpret and analyze the evidence we are gathering.

**ACKNOWLEDGMENTS AND FUNDING DISCLOSURE**

The authors gratefully acknowledge the support of the National Science Foundation, "EAGER CNS: Computational Propaganda and the Production/Detection of Bots," BIGDATA-1450193, 2014-16, Philip N. Howard, Principal Investigator. The research leading to these results has received funding from the European Research Council under the European Union's Consolidator Award #648311. Project activities were approved by the University of Washington Human Subjects Committee, approval #48103-EG. Any opinions, findings, and conclusions or recommendations expressed in this material are those of the author(s) and do not necessarily reflect the views of either the National Science Foundation or European Research Council.